\documentclass[fleqn,usenatbib]{mnras}

\usepackage{newtxtext,newtxmath}
\usepackage[T1]{fontenc}
\usepackage{ae,aecompl}
\usepackage{pdflscape}
\usepackage{afterpage}

\usepackage{graphicx}	% Including figure files

\usepackage{amsmath}	% Advanced maths commands
\usepackage{amssymb}	% Extra maths symbols
\usepackage{subfig}
\usepackage{booktabs}
\usepackage{afterpage}
\usepackage{float}
\usepackage{tablefootnote}
\usepackage{hyperref}
\usepackage{threeparttable}
\newcommand{\mysim}{\mathord{\sim}}
\newcommand{\myapprox}{\mathord{\approx}}

\usepackage{xcolor}

\usepackage[normalem]{ulem}

\defcitealias{Bonafede2010}{B10}
\newcommand{\refBA}{{\citetalias{Bonafede2010}} }
\newcommand{\refBAs}{{\citetalias{Bonafede2010}}}
\defcitealias{Bonafede2022}{B22}
\newcommand{\refBB}{{\citetalias{Bonafede2022}} }
\newcommand{\refBBs}{{\citetalias{Bonafede2022}}}

\title[Coma nonthermal]{Coma cluster $\gamma$-ray and radio emission is consistent with a secondary electron origin for the radio halo}

\author[Kushnir et al.]{
	Doron Kushnir$^{1}$, Uri Keshet$^{2}$ and Eli Waxman$^{1}$
	\thanks{E-mail: doron.kushnir@weizmann.ac.il}
	\\
	$^{1}$Dept. of Particle Phys. \& Astrophys., Weizmann Institute of Science, Rehovot 76100, Israel\\
    $^{2}$Physics Department, Ben-Gurion University of the Negev, POB 653, Be'er-Sheva 84105, Israel\\
}

\date{Accepted XXX. Received YYY; in original form ZZZ}

\pubyear{2024}

\begin{document}
\label{firstpage}
\pagerange{\pageref{firstpage}--\pageref{lastpage}}
\maketitle

% Abstract of the paper
\begin{abstract}
Observations of diffuse, non-thermal radio emission spanning several megaparsecs have been documented in over 100 galaxy clusters. This emission, classified as giant radio halos (GHs), mini halos, and radio relics based mainly on their location and morphology, is interpreted as synchrotron radiation and implies the presence of relativistic electrons and magnetic fields in the intra-cluster medium (ICM). GHs were initially thought to be generated by secondary electrons resulting from inelastic $p+p\rightarrow X+\pi^{\pm}$ collisions. However, recent literature has leaned towards primary-electron turbulent (re)acceleration models, partly due to claimed upper limits on the $\gamma$-ray emission from $\pi^0$ decay. We demonstrate that the observed GH and $\gamma$-ray flux in the Coma cluster are consistent with a secondary origin for the GH across a broad range of magnetic field values. Although the constraints on magnetic field configuration are not stringent, they align well with previous estimates for Coma. Within this magnetic field range, the energy density of cosmic-ray protons (CRp) constitutes a few percent to tens of percent of the ICM energy density, as predicted and observed for a sample of radio-emitting galaxy clusters. Notably, we detect a rise in the ratio of CRp to ICM energy densities towards the outer regions of the cluster. This phenomenon was anticipated to arise from either adiabatic compression of CRp accelerated by accretion shocks or, more likely, from strong CRp diffusion.
\end{abstract}

% Select between one and six entries from the list of approved keywords.
% Don't make up new ones.

%%%%%%%%%%%%%%%%%%%%%%%%%%%%%%%%%%%%%%%%%%%%%%%%%%

%%%%%%%%%%%%%%%%% BODY PAPER %%%%%%%%%%%%%%%%%%

\section{introduction}
\label{sec:introduction}
Diffuse, non-thermal radio emission on scales as large as a few Mpc has been observed in more than $100$ clusters of galaxies \citep[see][for a recent review]{VanWeeren2019}. The radio emission is interpreted as synchrotron radiation, suggesting that relativistic electrons and magnetic fields exist in the intra-cluster medium (ICM). The emission is classified as giant radio halos (GHs), mini halos (MHs), and radio relics, depending mainly on their location and morphology. Several models for synchrotron emission in halos have been presented in the literature. These models differ in the assumptions regarding the origin of the emitting electrons. In some models, the emitting electrons are secondary electrons and positrons that were generated by $p-p$ interactions of a cosmic-ray proton (CRp) population with the ICM \citep[e.g.][]{Dennison1980,Blasi1999} while in others, the emitting electrons are reaccelerated by turbulence from a preexisting population of nonthermal seeds in the ICM \citep[secondary or otherwise, e.g.][]{Brunetti2001,GittiEtAl02}.

Secondary (i.e. hadronic) models explain the diffuse radio emission from the ICM more naturally than primary (leptonic) models, not only in GHs \citep{Kushnir2009b}, MHs \citep{KeshetLoeb10}, and even relics \citep{Keshet10}, but also in the transitional, hybrid stages of these sources  \citep{Keshet10,Keshet23}, which in primary models would require fine-tuning different acceleration mechanisms to match each other. Fundamentally, while the presence of extended CR ion populations in clusters is thought to be both unavoidable and at the level needed to produce the observed secondary-electron radio sources, primary models invoke non-standard processes of electron (re)acceleration in turbulence or weak shocks, which are neither well established theoretically, nor confirmed observationally elsewhere, nor plausible in the presence of the strong diffusion inferred in the ICM \citep{Keshet23}.

The secondary nature of these diffuse sources is affirmed by accumulated observations, including morphological, energetic, and spectral evidence tying GHs, MHs, relics, and transients such as halo--relic bridges as different manifestations of the same underlying mechanism; the nearly universal, $-1.2\lesssim\alpha\lesssim -1.0$ spectral index of integrated sources (except at transitional stages), including weak-shock relics (which in primary models should be softer); sufficient acceleration of CRp in virial shocks established by $\gamma$-ray to radio \citep{KeshetEtAl12_Coma, ReissKeshet18, HouEtAl22} emission detected from their CR electron counterparts; and radio--X-ray \citep{Kushnir2009b} and other correlations between coincident signals \citep[see][and references therein]{Keshet23}.

Arguments raised against GHs as secondary sources target their spatial and spectral properties \citep[\emph{e.g.,}][]{BrunettiJones14}, but such criticisms tend to invoke oversimplified hadronic models. Some GHs, such as in A521, show a very soft spectrum but are likely transient sources associated with a very recent merger, in which rapid magnetic growth \citep{Keshet10} combined with strong CR diffusion \citep{Keshet23} softens the spectrum. Even evolved GHs such as in Coma, with a flat integrated spectrum, can show regions of very soft spectra in particular in their peripheries, but such softening is naturally explained by the strong CR electron diffusion inferred observationally \citep{Keshet23}.

The allegedly strongest and most decisive argument invoked against the secondary model in GHs is that the CRp population required to produce the radio emission from the Coma cluster emits a $\gamma$-ray flux through $p-p$ collisions that is larger than observed or constrained by \textit{Fermi}-LAT \citep{BrunettiZimmer2017,VanWeeren2019,Adam2021}.

In this paper, we show that the GH and $\gamma$-ray flux of the Coma cluster are fully consistent with secondaries from the same CRp population. The relevant radio and $\gamma$-ray observations for our analysis are described in Section~\ref{sec:observations}. We focus on the $144\,\rm{MHz}$ LOFAR observations \citep[][hereafter \refBBs]{Bonafede2022} and on \textit{Fermi}-LAT observations of the Coma cluster. We present in Section~\ref{sec:consistency} an order of magnitude assessment utilizing the ratio of radio to $\gamma$-ray flux, demonstrating the consistency of the observations with a secondary source for the GH. A straightforward model for the CRp population and magnetic field is presented in Section~\ref{sec:model}. In Section~\ref{sec:results}, we calculate the CRp population required to reproduce the observed GH emission for a given magnetic field. We then calculate the $\gamma$-ray flux from this population and compare it to the \textit{Fermi}-LAT observations. We find that the secondary origin of the Coma GH is consistent with observations over a large range of magnetic field values. Over this range, the CRp energy density is a few percent to tens of percent of the ICM energy density. We find that the ratio between the CRp and the ICM energy densities increases towards the cluster's outskirts. This observation was anticipated to arise from either adiabatic compression of accretion-shock accelerated CRp \citep{Kushnir2009} or, more favorably, as the result of strong CRp diffusion \citep{Keshet10,Keshet23}, with unconstrained accelerators. In Section~\ref{sec:discussion}, we discuss the implications of our findings and show, in particular, that an exceedingly high magnetic field, regarded as necessary for the hadronic model in prior investigations, stems from the steep CRp spectrum assumed in these studies.

\section{Observations}
\label{sec:observations}

In this section, we describe the relevant radio (Section~\ref{sec:radio}) and $\gamma$-ray (Section~\ref{sec:gamma}) observations used in our analysis.

\subsection{Radio}
\label{sec:radio}

\begin{figure*}
	\includegraphics[width=0.8\textwidth]{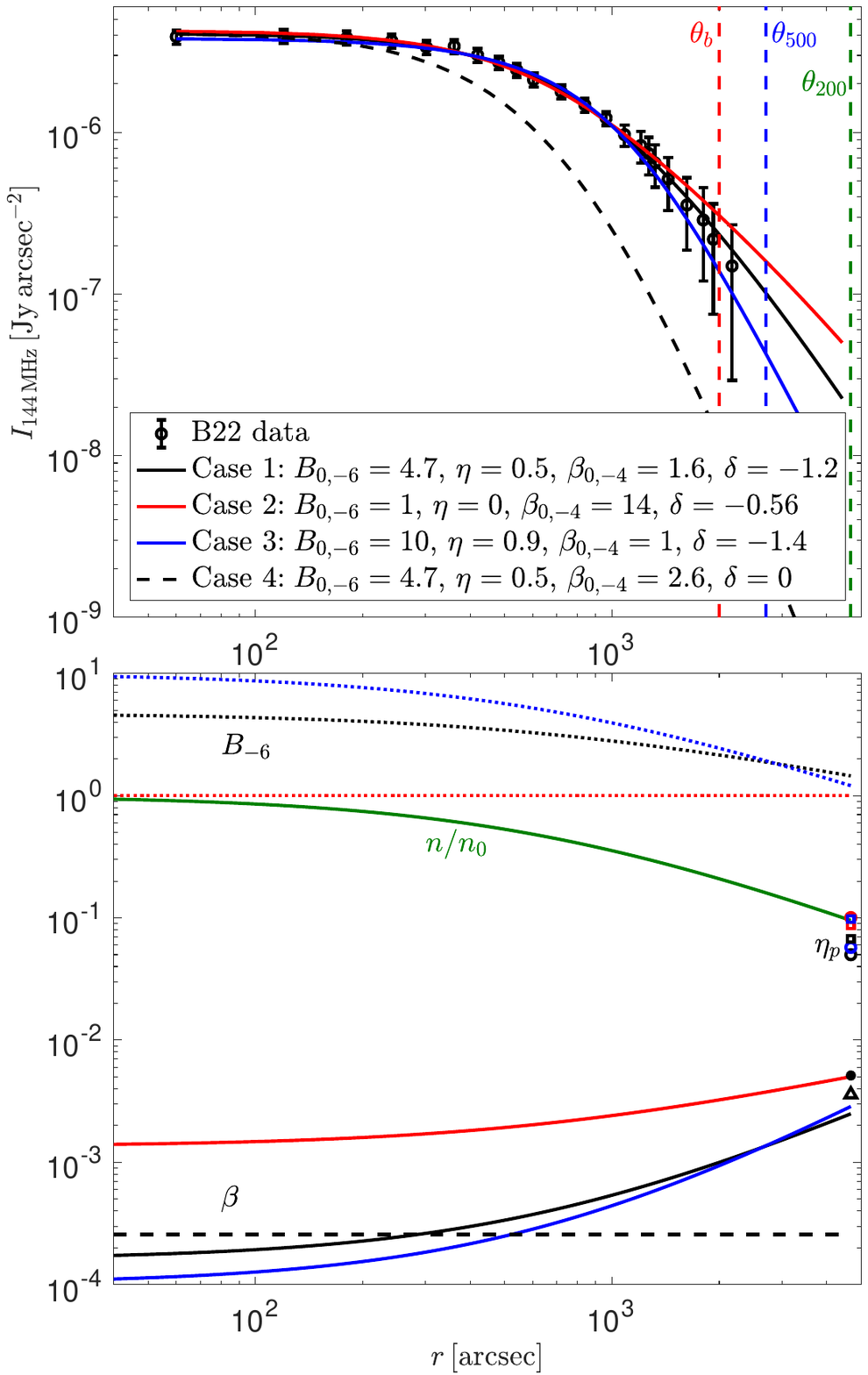}
	\caption{Results of the fitting as a function of the distance from the cluster center for several example values of $B_0$ and $\eta$. Upper panel: Radio surface brightness at $144\,\rm{MHz}$. Black symbols represent the mean radial profile of \refBB, with error bars denoting the scatter between the three sectors averaged for the mean profile. Solid lines and a dashed black line depict the results for cases 1-4 (refer to the text for details). Dashed vertical lines denote $\theta_b$, $\theta_{500}$, and $\theta_{200}$ (as described in the text), highlighted in red, blue, and green, respectively. Lower panel: The assumed magnetic field profile (illustrated with dotted lines) and the calibrated $\beta$ profiles (displayed with solid lines and a dashed black line) for each case (using the same colors and line types as in the upper panel). Circles represent the extrapolated post-shock thermal plasma energy deposited in CRp ($\eta_p$), with open circles corresponding to cases 1-3 (with matching colors) and the filled circle corresponding to case 4. Squares and a triangle denote the ratio between the total energy of the CRp component (integrated up to $\theta_{200}$) and the thermal energy of the gas, $E_{nt}/E_{th}$. Squares correspond to cases 1-3 (with matching colors), while the triangle corresponds to case 4. A green line illustrates the assumed electron density of the ICM.
}
	\label{fig:Examples}
\end{figure*}

The \refBB $144\,\rm{MHz}$ observations of the Coma cluster include a radio halo, a radio relic, and a bridge connecting the two. \refBB provides the radial radio brightness profiles of the GH in three sectors (excluding the SW sector, which harbors the bridge region); see their figure 12. The profiles are computed in elliptical annuli, with a major-to-minor axis ratio of $\mysim1.3$. In what follows, we treat the radial profiles as circular and average them, which may lead at most to a few tens of percent error, not impacting our main conclusions. Moreover, we note that for a slightly higher frequency, $342\,\rm{MHz}$, the major-to-minor axis ratio decreases to $\mysim1.1$, approaching spherical symmetry. We defer the construction of a more detailed model to reflect the deviations of the radio surface brightness from a circular shape to future work. The mean radial profile is presented with black symbols in the top panel of Figure~\ref{fig:Examples}. The error bars reflect the scatter between the three sectors. The spectral index computed between $144\,\rm{MHz}$ and $342\,\rm{MHz}$ is consistent with $\alpha=-1$, up to $\theta_{b}\approx2000''$ (red dashed line), within which $>95\%$ of the surface brightness resides. The spectral index becomes steeper at larger radii.

It is worth noting for future reference that the total radio flux at $144\,\rm{MHz}$ emitted by the cluster within an angular radius of $\theta_{500}\approx0.75^{\circ}$ (indicated by the blue dashed line in the upper panel of Figure~\ref{fig:Examples}) amounts to $f_r\equiv2\pi\int \nu I_\nu rdr\approx1.7\times10^{-14}\,\rm{erg\,s^{-1}\,cm^{-2}}$. This estimation encompasses a slight extrapolation of around $5\%$ from the last radio data point to $\theta_{500}$, and it remains unaffected by the spatial modeling of the brightness profile.

\subsection{$\gamma$-rays}
\label{sec:gamma}

The $\gamma$-ray flux from the Coma cluster has been analyzed in a few works. \citet{Xi2018} claimed a detection of $\mysim2.5\times10^{-9}\,\rm{ph}\,\rm{s}^{-1}\,\rm{cm}^{-2}$ diffuse emission within $1.23^{\circ}\approx\theta_{200}$ (green dashed line in the upper panel of Figure~\ref{fig:Examples}) in the energy range $[200\,\rm{MeV}-300\,\rm{GeV}]$ (see, e.g., their disc+p1 model). \citet{Adam2021} claimed a detection of $\mysim10^{-9}\,\rm{ph}\,\rm{s}^{-1}\,\rm{cm}^{-2}$ within $\theta_{500}$ in the same energy range. The exact flux depends on the underlying models for the background and the signal angular distribution. We include this by taking a $\mysim3$ uncertainty factor on the measured flux \citep[see, e.g., table 1 of][]{Adam2021}. \citet{Baghmanyan2021} reported a [$200$ MeV--$300$ GeV] flux of $(2.41\pm0.45)\times 10^{-9}\rm{\,ph\,s}^{-1}\rm{cm}^{-2}$ for a uniform disc model, probably adopting their $0.82_{-0.05}^{+0.10}$ degree $68\%$ containment radius. The spectral index of the emission is consistent with a flat spectrum, although difficult to measure and model-dependent: \citet{Xi2018} obtained $2.2\lesssim\Gamma\lesssim3.2$, \citet{Adam2021} obtained $\Gamma=2.45\pm0.19$, and \citet{Baghmanyan2021} obtained $\Gamma=2.23\pm0.11$. Skeptical readers may use $3\times10^{-9}\,\rm{ph}\,\rm{s}^{-1}\,\rm{cm}^{-2}$ as an upper limit for the $\gamma$-ray flux from the cluster.

\section{The consistency of the observations with a secondary origin for the GH}
\label{sec:consistency}

Prior to employing a straightforward model in Section~\ref{sec:model} to demonstrate the consistency of the observations with a secondary source for the GH, it proves beneficial to conduct an order of magnitude assessment utilizing the ratio of radio to $\gamma$-ray flux. We adopt a flat CRp spectrum ($dn/d\varepsilon\propto\varepsilon^{-2}$; denoted as $\delta_p=-2$), which generates $\gamma$-rays and secondary particles (electrons and positrons) through $p-p$ collisions with the nuclei of the ICM. Our choice of the CRp spectral index, where the energy density per logarithmic energy interval remains constant with energy as expected for particles accelerated at high Mach number shocks \citep{Blandford1987}, aligns with the observed spectral indices of both the radio halo and $\gamma$-ray emissions (refer below). This flat spectrum is also consistent with the spectrum observed from virial shocks in stacked $\gamma$-ray \citep{ReissKeshet18} and radio \citep{HouEtAl22} clusters and in the Coma cluster \citep{KeshetReiss18}, as well as with the aforementioned universal $-1.2\lesssim\alpha\lesssim-1.0$ of integrated GHs, MHs, and relics. We assume a scenario where the distribution of radio-emitting secondaries is in a steady state, wherein within the relevant energy ranges, the generated secondaries lose all their energy to synchrotron radiation and inverse-Compton scattering of Cosmic Microwave Background (CMB) photons.

In the context of a flat CRp spectrum, the energy density per logarithmic energy interval remains invariant with energy, hence the $p-p$ $\gamma$-ray luminosity per logarithmic photon energy bin exhibits an approximately energy-independent behavior, $\Gamma\equiv 1-d\ln f_{\varepsilon}/d\ln {\varepsilon}\approx-\delta_p=2$, where $f_{\varepsilon}$ is the energy flux per unit energy, consistent with the observations (note that constraining the $\gamma$-ray spectrum in an individual cluster is challenging, as discussed in Section~\ref{sec:gamma}). Likewise, the synchrotron emissivity per logarithmic frequency interval is directly proportional to the energy production rate of secondaries per logarithmic secondary energy interval, also maintaining an energy-independent spectrum, resulting in $\alpha\approx\delta_p/2=-1$, consistent with the radio observations within the hundreds of MHz frequency range (refer to the right panel of \refBB Figure 9).

The ratio of the radio to the $\gamma$-ray flux is given to a good approximation by \citep{Katz2008,Kushnir2009}
 \begin{equation}\label{eq:simple ratio}
 \frac{\nu I_{\nu}}{\varepsilon f_{\varepsilon}}\approx\frac{1}{4}\frac{B^{2}}{B^{2}+B^{2}_{\rm{CMB}}}\approx0.1,
 \end{equation}
where $B_{\rm{CMB}}\approx3\mu \textrm{G}$ represents the equivalent CMB energy density magnetic field, and we employed $B=3\,\mu\textrm{G}$ for the last equality. Utilizing $f_r\approx1.7\times10^{-14}\,\rm{erg\,s^{-1}\,cm^{-2}}$ from Section~\ref{sec:radio} and the $\gamma$-ray flux  $\mysim3\times10^{-10}-3\times10^{-9}\,\rm{ph}\,\rm{s}^{-1}\,\rm{cm}^{-2}$ from Section~\ref{sec:gamma} (within $\theta_{500}$), yields (with $\varepsilon=200\,\rm{MeV}$):
\begin{equation}
\frac{\nu I_{\nu}}{\varepsilon f_{\varepsilon}}\approx0.02-0.2,
\end{equation}
consistent with Equation~\eqref{eq:simple ratio}. This consistency persists with the more detailed model outlined in Section~\ref{sec:model}.

The large magnetic field, regarded as necessary for the hadronic model in prior investigations, stems from the steep CRp spectrum assumed in these studies (refer to Section~\ref{sec:discussion} for a detailed discussion).  For now, note that the frequency at which an electron emits most of its synchrotron power is dictated by $\nu=\nu_0\gamma^2$, where $\nu_0=3eB/(4\pi m_e c)$ and $\gamma$ represents the Lorentz factor of the electron. Consequently:
\begin{equation}
    \gamma\approx6\times10^3\left(\frac{\nu}{144\,\rm{MHz}}\right)^{1/2}B_{-6}^{-1/2},
\end{equation}
indicating an electron energy of approximately $\mysim3\,$GeV. Since the energy of secondary electrons from a $p-p$ collision is roughly half of the $\gamma$-ray energy (for a given CRp energy), the parent CRp of the $\mysim3\,$GeV electrons produce $\mysim6\,$GeV $\gamma$-rays, surpassing the $200\,\rm{MeV}$ energy of the $\gamma$-rays which contribute most to the flux. This discrepancy signifies that the parent CRp of the electrons carries more than $10$ times the energy of the parent CRp of the $\gamma$-rays, with implications discussed in Section~\ref{sec:discussion}. Finally, note that the cooling time of these electrons, ranging from  $\mysim0.1-0.4\,\rm{Gyr}$ for $B_{-6}=0.1-10$, is shorter than the dynamical time of the cluster, allowing the application of the steady-state assumption.

\section{Model}
\label{sec:model}

In our simple model, the ICM contains a population of CRp with a power law distribution $\varepsilon^{2}dn/d\varepsilon=\beta3nT/2$, where $n$ and $T$ are the number density and (constant) temperature of the ICM, respectively. We assume that $\beta$, the ratio of the CRp energy density (within a logarithmic CRp energy interval) and the thermal energy of the ICM, scales with the ICM number density as $\beta=\beta_0(n/n_0)^{\delta}$, where subscript $0$ denotes values at the center of the cluster. Since CRp produced by the accretion shock are later affected mainly by adiabatic expansion and compression, the difference between the adiabatic indices of the relativistic CRp ($4/3$) and nonrelativistic thermal plasma ($5/3$) implies $\delta=-1/3$ \citep{Kushnir2009}. More detailed modeling suggests that typically $\delta<-1/3$ because of the effect of weak merger shocks \citep{Kushnir2009}. If CRp diffusion is sufficiently strong, we expect a constant CRp density, $\delta=-1$, regardless of the CRp acceleration sites (as long as the acceleration is at high Mach number shocks to produce a flat spectrum),

The electron density of the ICM is assumed to follow a $\beta_m$-model \citep{Cavaliere1978}, $n=n_0(1+(r/r_c)^2)^{-3\beta_m/2}$, and we adopt the values $n_0=3.36\times10^{-3}\,\rm{cm}^{-3}$, $r_c=310\,\rm{kpc}$, $\beta_m=0.75$, used by \citet{Adam2021} (see green line in the lower panel of Figure~\ref{fig:Examples}). We further assume a constant temperature, $T=8.25\,\textrm{keV}$, for the ICM \citep{Reiprich2002}.

The magnetic field strength profile of the Coma cluster was inferred by \citet[][hereafter \refBAs]{Bonafede2010} using Faraday rotation measures to be $B(n)=B_0(n/n_0)^{\eta}$ with the best-fit values of $B_{0,-6}=4.7$ ($B_{-6}=B/\mu\rm{G}$) and $\eta=0.5$. Here we adopt the parametric form of \refBAs, but due to possible biases in the analysis of \refBA \citep[see, e.g., discussion in][]{BrunettiZimmer2017}, we allow a large range of $B_0$ and $\eta$ values.

We use the parametrization of \citet{Kamae2006} for the secondary spectrum of $p-p$ interactions, and the exact formula for synchrotron emission given by \citet{Blumenthal1970} to calculate the $144\,\rm{MHz}$ surface brightness and the $[200\,\rm{MeV}-300\,\rm{GeV}]$ $\gamma$-ray flux.

\section{Results}
\label{sec:results}

For each combination of $B_0$ and $\eta$, we find the values of $\beta_0$ and $\delta$ that best fit the observed $144\,\rm{MHz}$ surface brightness. For example, our results for $B_{0,-6}=4.7$ and $\eta=0.5$ (dotted black line at the lower panel of Figure~\ref{fig:Examples}) are presented in the top panel of Figure~\ref{fig:Examples} (black solid line; Case 1). We find $\beta_{0,-4}=1.6$ (defining $\beta_{0,-4}=\beta_0/10^{-4}$) and $\delta=-1.2$, which nicely agree with the observed surface brightness. The value of $\beta_0$ is close to the value $2\times10^{-4}$ calibrated to Coma by \citet{Kushnir2009}, to a sample of radio-emitting galaxy clusters by \citet[][based on the correlation between the radio luminosity and the thermal X-ray luminosity]{Kushnir2009b}, and to a sample of GHs and MHs by \citet{KeshetLoeb10}. The value of the calibrated $\delta$ satisfies the expectation that $\delta<-1/3$ (the radial profile of $\beta$ is presented as a solid black line at the lower panel of Figure~\ref{fig:Examples}). Interestingly, the value of $\delta$ is close to the strong diffusion expectation ($-1$), suggesting this limit is achieved, regardless of the CRp acceleration sites. The obtained $\gamma$-ray flux for the calibrated $\beta_0$ and $\delta$ values is $\myapprox0.85(1.45)\times10^{-9}\,\rm{ph}\,\rm{s}^{-1}\,\rm{cm}^{-2}$ within $\theta_{500}(\theta_{200})$, in an agreement with the observations of \citet{Adam2021} \citep{Xi2018} and slightly below the observation of \citet{Baghmanyan2021}. A significant fraction of the $\gamma$-ray flux, $\myapprox77(45)\%$, is contained within the angular distance for which the radio observations directly constrain $\beta$, such that the uncertainty due to the extrapolation to $\theta_{500}(\theta_{200})$ is not large.

We may also extrapolate $\beta$ to $\theta_{200}\approx1.3^{\circ}$ (dashed green line in the top panel of Figure~\ref{fig:Examples}), where the accretion shock is predicted to reside (but see discussion in Section~\ref{sec:discussion}). We can then multiply the extrapolated $\beta$ by the number of logarithmic intervals that span the CRp distribution \citep[$\myapprox20$; see, e.g., ][]{Kushnir2009}  to find the fraction $\eta_p\approx3.7\times10^{-2}$ of the post-shock thermal plasma energy deposited in CRp (the black disc in the bottom panel of Figure~\ref{fig:Examples}). The ratio between the total energy of the CRp component (integrated up to $\theta_{200}$) and the thermal energy of the gas is $\myapprox0.005$ per logarithmic CRp energy interval, leading to a ratio of $E_{nt}/E_{th}\approx0.07-0.1$ between these two components, depending on the number of logarithmic intervals that the CRp distribution spans ($\myapprox14(20)$ for maximal energy of $10^{6(9)}\,\rm{GeV}$ relevant for supernova remnants (galaxy cluster accretion shocks)). This estimate is plotted with a black square in the bottom panel of Figure~\ref{fig:Examples}.

\begin{figure*}
	\includegraphics[width=\textwidth]{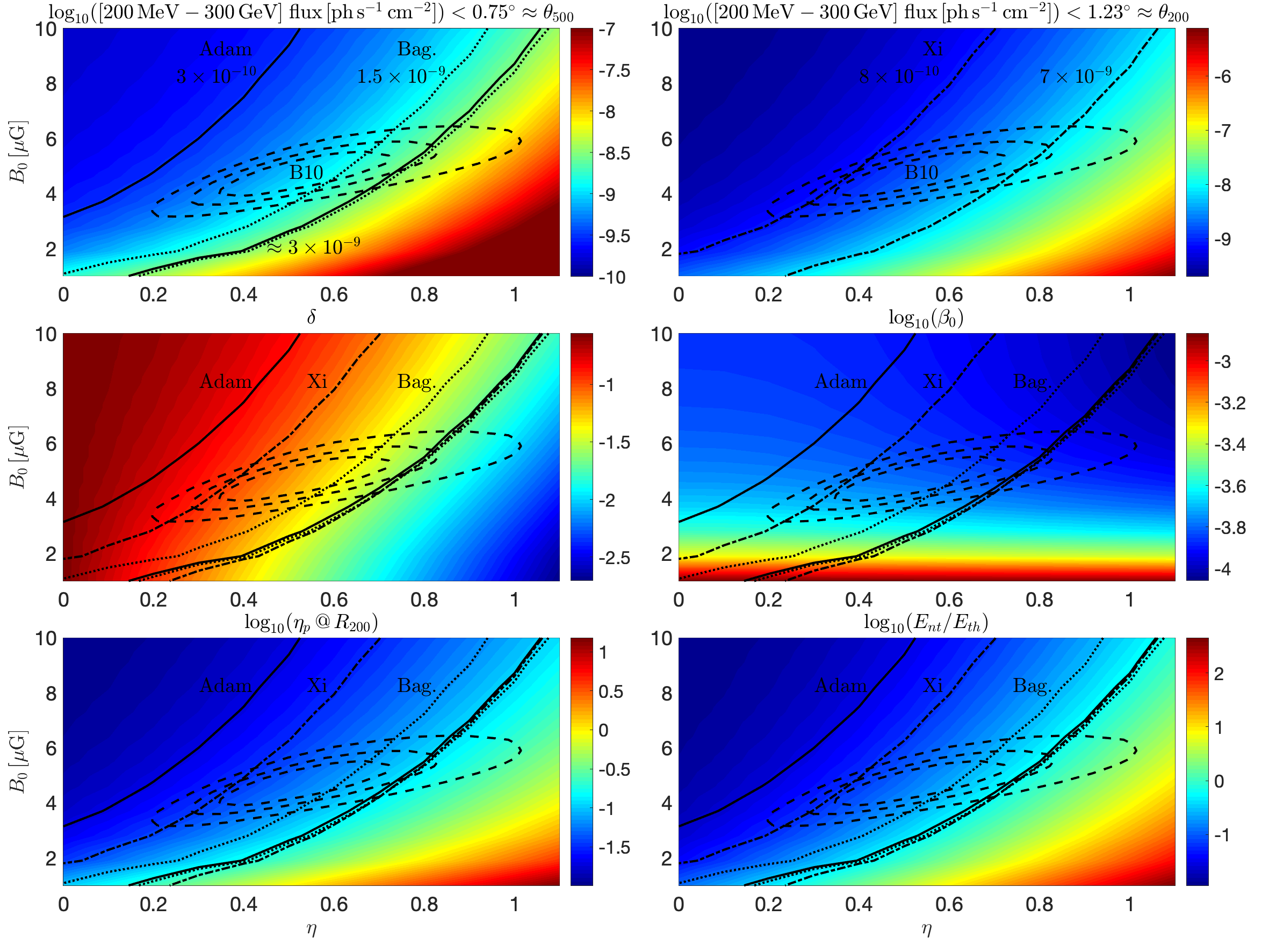}
	\caption{The outcomes of the analysis for each pairing of $B_{0}$ and $\eta$ within the surveyed range. The enclosed dashed contours delineate the best-fit values of \refBA (same $1-3\sigma$ contours in all panels). Top-left (top-right) panels: The obtained $[200\,\rm{MeV}-300\,\rm{GeV}]$ $\gamma$-ray flux within angular diameter of $0.75^{\circ}\approx\theta_{500}$ ($1.23^{\circ}\approx\theta_{200}$) as inspected by \citet{Adam2021} and \citet{Baghmanyan2021} \citep[][]{Xi2018}. The constrained $\gamma$-ray flux of \citet{Adam2021} and \citet{Baghmanyan2021} lies between the two solid and dotted black lines, respectively \citep[we plot the 2$\sigma$ constraint of][]{Baghmanyan2021}. The constrained $\gamma$-ray flux of \citet{Xi2018} lies between the dot-dashed black lines. Subsequent panels depict colormaps of the calibrated $B_0$, $\log_{10}(\beta_0)$, $\log_{10}(\eta_p)$ at $R_{200}$, and $\log_{10}(E_{nt}/E_{th})$. The black lines correspond to the lines in the upper panels ("Adam" for \citet{Adam2021}, "Xi" for \citet{Xi2018}, and "Bag." for \citet{Baghmanyan2021}). The values within the calibrated parameter ranges are all reasonable, precluding stronger constraints on combinations of $B_0$ and $\eta$.
}
	\label{fig:FullDist}
\end{figure*}

We next allow $B_{0,-6}$ to vary in the range $1-10$ and $\eta$ to vary in the range $0-1.1$. We repeat our analysis for each combination of $B_{0}$ and $\eta$, and the results are presented in Figure~\ref{fig:FullDist}. As can be seen in the top-left (top-right) panel of Figure~\ref{fig:FullDist}, the $\gamma$-ray observations of \citet{Adam2021} and \citet{Baghmanyan2021} \citep{Xi2018} are consistent with a wide range of $B_0$ and $\eta$ values (all combinations with $\gamma$-ray flux within the two black lines; we plot the 2$\sigma$ constraints of \citet{Baghmanyan2021}). Specifically, the constraints for $B_0$ and $\eta$ of \refBA are consistent with the observations. Two extreme cases of $B_0$ and $\eta$ values, $B_{0,-6}=1$, $\eta=0$ and $B_{0,-6}=10$, $\eta=0.9$ (Case 2 and Case 3, respectively), which are consistent with the observations, are presented in Figure~\ref{fig:Examples} (with red and blue colors, respectively). While the agreement with the radio surface brightness at the largest radial bins is slightly worse than the agreement in Case 1, these cases cannot be ruled out based on the radio observations. The ranges of the calibrated $\beta_{0,-4}$ and $\delta$ is $[0.94,13.5]$ and $[-1.7,-0.56]$ (middle-right and middle-left panels of Figure~\ref{fig:FullDist}, respectively) for all models with consistent $\gamma$-ray flux. For the same models, the obtained range of $\eta_p$ is $[0.02,0.25]$ (for $20$ logarithmic intervals of the CRp energy distribution) and of $E_{nt}/E_{th}$ is $[0.02,0.6]$ (for $14$ logarithmic intervals of the CRp energy distribution). The values within the ranges of the calibrated parameters are all reasonable, not allowing stronger constraints on combinations of $B_0$ and $\eta$.

The data allows us to rule out the case of density-independent $\beta$ (i.e. $\delta=0$). Forcing $\delta=0$ for $B_{0,-6}=4.7$ and $\eta=0.5$ leads to a radial surface brightness profile that is too concentrated (dashed black line in the top panel of Figure~\ref{fig:Examples}, Case 4). The same argument holds for all $\delta>0$ values.

\section{Discussion}
\label{sec:discussion}

We showed that the GH and $\gamma$-ray flux of the Coma cluster are fully consistent with a secondary origin for the GH over a large range of magnetic field values. Although the constraints on the magnetic field configuration are not stringent, they align well with previous estimates for Coma ({\refBAs}). Within this magnetic field range, the energy density of CRp constitutes a few percent to tens of percent of the ICM energy density, as predicted and observed for a sample of radio-emitting galaxy clusters \citep{Kushnir2009b}. We find that the ratio between the CRp and the ICM energy densities increases towards the cluster's outskirts. This observation was anticipated to arise from either adiabatic compression of accretion-shock accelerated CRp \citep{Kushnir2009} or, more favorably, as the result of strong CRp diffusion \citep{Keshet23}, allowing also for other possibilities for the acceleration sites.

Our analysis assumed a spherical symmetry for the cluster, probably introducing at most a few tens of percent error. We defer the construction of a more detailed model to reflect the deviations of the radio surface brightness from a circular shape to future work.

The recent radio and $\gamma$-ray observations of the Coma cluster are found to be entirely consistent with a hadronic origin, contrasting with previous claims \citep{BrunettiZimmer2017,VanWeeren2019,Adam2021}. In these earlier studies, a substantially large magnetic field was deemed necessary for the hadronic model, implying that a significant portion of the Coma cluster's energy was attributed to the magnetic field. However, our analysis reveals a different scenario, where the best-fit values of \refBA align well with our hadronic model. The discrepancy arises from the differing assumptions regarding the CRp spectrum. Previous works assumed a steep CRp spectrum ($\delta_p<-2$), while we adopted a flat CRp spectrum ($\delta_p=-2$). The steep spectrum in previous studies was motivated by the purported steep spectrum of the GH (note that the $\gamma$-ray spectrum is difficult to constrain, see Section~\ref{sec:gamma}). For instance, \citet{BrunettiZimmer2017} claimed $\alpha=-1.22\pm0.04$, resulting in $\delta_p\approx2\alpha=-2.44$. However, the GH spectral index was determined by fitting a broad range of radio frequencies, including high frequencies where the parent CRp's energy is too high to contribute to the observed $\gamma$-ray flux (see below). When limiting the radio frequency range to hundreds of MHz, a slope of $\alpha\simeq-1$ emerges (see the right panel of \refBB Figure 9; the observed spectral softening at the edge of the GH is expected for secondary electrons diffusing outward \citep[see][]{Keshet23}), rendering the index assumed by \citet{BrunettiZimmer2017} untenable. Other analyses adopted even softer CRp spectra, such as $\delta_p<-2.7$ in \citet{Adam2021}, largely based on the poorly determined $\gamma$-ray spectrum.

\begin{figure}
	\includegraphics[width=0.5\textwidth]{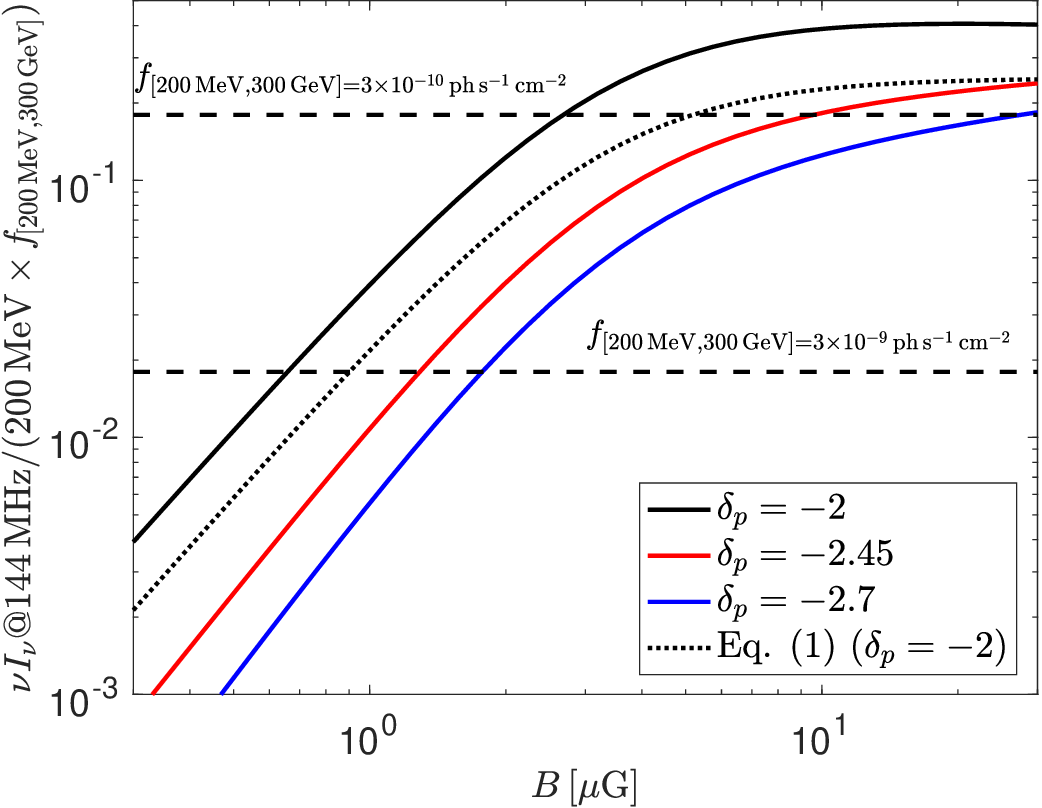}
	\caption{The relationship between the radio to $\gamma$-ray flux ratio and the magnetic field. Solid lines represent the calculated ratios for different indexes of the CRp spectra ($\delta_p=-2,-2.45,-2.7$ in black, red, and blue, respectively). The dotted line shows the simple estimate from Equation~\eqref{eq:simple ratio} pertinent for $\delta_p\approx-2$, exhibiting a deviation by less than a factor of $2$ from the more precise calculation. The observed ratio falls within the two dashed lines (within $\theta_{500}$, see Section~\ref{sec:consistency}). The calculation for $\delta_p=-2$ is consistent with observations for $0.5<B_{-6}<3$. However, considering that the parent CRp of electrons carries more than $10$ times the energy of the parent CRp of $\gamma$-rays (see Section~\ref{sec:consistency}), a steep spectrum increases the energy density of the parent CRp of electrons relative to that of $\gamma$-rays, necessitating larger magnetic fields to maintain a fixed ratio of radio to $\gamma$-ray flux.}
	\label{fig:Deltap}
\end{figure}

Assuming a steep spectrum for the CRp led those studies to artificial inflation of the energy density of low-energy CRp, resulting in an overproduction of $\gamma$-rays and effectively increasing the required magnetic field. This effect is illustrated in Figure~\ref{fig:Deltap}, where we depict the ratio of the radio to $\gamma$-ray flux as:
\begin{equation}
 \frac{\nu I_{\nu}}{200\,\textrm{MeV} f_{[200\,\textrm{MeV},200\,\textrm{GeV}]}},
\end{equation}
plotted against the magnetic field. This ratio is observed to fall within the range $0.02-0.2$ (within $\theta_{500}$, see Section~\ref{sec:consistency}). The figure displays the obtained ratio for calculations assuming different indexes for the CRp spectra, utilizing the parametrization of \citet{Kamae2006} for the secondary spectrum of $p-p$ interactions and the exact formula for synchrotron emission by \citet{Blumenthal1970} to compute the radio and $\gamma$-ray fluxes. Additionally, the simple estimate from Equation~\eqref{eq:simple ratio} pertinent for $\delta_p\approx-2$ is included, exhibiting a deviation by less than a factor of $2$ from the more precise calculation. As depicted in the figure, the calculation for $\delta_p=-2$ aligns with observations for $0.5<B_{-6}<3$, consistent with earlier sections. However, given that the parent CRp of the electrons carries more than $10$ times the energy of the parent CRp of the $\gamma$-rays (refer to Section~\ref{sec:consistency}), a steep spectrum augments the energy density of the parent CRp of the electrons compared to the energy density of the parent CRp of the $\gamma$-rays, necessitating larger magnetic fields to uphold a fixed ratio of radio to $\gamma$-ray flux, as evidenced in Figure~\ref{fig:Deltap} for $\delta_p=-2.45,-2.7$. As demonstrated here, for $\delta_p=-2$, which is fully consistent with the GH spectral index of \refBBs, the magnetic field constraints align perfectly with the best-fit values of {\refBAs}. It's worth noting that a slight deviation from $\delta_p=-2$ has a negligible impact on our findings, as the energy difference of the parent CRp is only about one order of magnitude.

It should be emphasized that previous studies that argued for inconsistency between hadronic halo interpretations and $\gamma$-ray observations could not show a substantial discrepancy exceeding the underlying systematic uncertainties, as evident from the variability in the results obtained by different authors and the arguments sketched above. For instance, \citet{Adam2021} argued that the secondary counterpart to their $\gamma$-ray model is too weak to explain the halo, necessitating some unknown primary electron acceleration. However, the discrepancy they report is only a factor of $\sim 4$, so the putative primary electron emission component should be comparable, i.e., only a factor $\sim 3$ stronger than the secondary component. Given the very different origins of secondary and putative primary electrons, such a coincidence is unlikely. On the contrary: given the substantial systematics and the challenges facing primary models, a radio--$\gamma$-ray discrepancy by a factor of a few in the hadronic picture should be considered a qualitative success of the model.

Recent evidence indicates that cluster virial shocks extend well beyond $\theta_{200}$, with strong stacked signatures found in the narrow range $[2.2,2.5]\theta_{500}$, or equivalently$[1.4,1.6]\theta_{200}$, and weak signatures of larger semi-major radii \citep{ReissKeshet18, HouEtAl22, IlaniEtAl24, IlaniEtAl24b}. We avoid extrapolating the $\beta$ profile beyond $\theta_{200}$, which would incur substantial systematic errors and may be unwarranted for the semi-minor axis in Coma \citep{KeshetEtAl12_Coma, KeshetReiss18}. Consequently, our extrapolated CRp fraction at the virial shock probably underestimates its true post-shock value, reaching tens of percent at $\gtrsim1.4\theta_{200}$. Hence, our results are consistent with strong cosmic-ray diffusion not only in terms of the $\delta\sim -1$ profile, but also, given the order-unity energy fraction typically anticipated in post-shock CRp, in our normalization of the CRp distribution; strong diffusion also accounts for the spectral softening at the edge of the GH \citep[as secondary electrons diffuse outward; see][]{Keshet23}, seen for example in the top panel of Figure \ref{fig:Examples}, and for additional observations such as the GH--relic bridge \citep{Keshet10, Keshet23}.

\section*{Acknowledgements}
We thank Boaz Katz for useful discussions. DK is supported by a research grant from The Abramson Family Center for Young Scientists and by the Minerva Stiftung.
UK is supported by ISF grant No. 2126/22.

\section*{Data availability}

All data used in this study is available through other publications.

%\bibliographystyle{mnras}
%\bibliography{bibliography}

\begin{thebibliography}{99}
\bibitem[\protect\citeauthoryear{Adam et al.}{2021}]{Adam2021} Adam R., Goksu H., Brown S., Rudnick L., Ferrari C., 2021, A\&A, 648, A60. doi:10.1051/0004-6361/202039660
\bibitem[\protect\citeauthoryear{Baghmanyan et al.}{2022}]{Baghmanyan2021} Baghmanyan V. , Zargaryan D., Aharonian F., Yang R., Casanova S., Mackey
J., 2022, MNRAS , 516, 562 doi:10.1093/mnras/stac2266
\bibitem[\protect\citeauthoryear{Blandford \& Eichler}{1987}]{Blandford1987} Blandford R., Eichler D., 1987, PhR, 154, 1. doi:10.1016/0370-1573(87)90134-7
\bibitem[\protect\citeauthoryear{Blasi \& Colafrancesco}{1999}]{Blasi1999} Blasi P., Colafrancesco S., 1999, APh, 12, 169. doi:10.1016/S0927-6505(99)00079-1
\bibitem[\protect\citeauthoryear{Blumenthal \& Gould}{1970}]{Blumenthal1970} Blumenthal G.~R., Gould R.~J., 1970, RvMP, 42, 237. doi:10.1103/RevModPhys.42.237
\bibitem[\protect\citeauthoryear{Bonafede et al.}{2010}]{Bonafede2010} Bonafede A., Feretti L., Murgia M., Govoni F., Giovannini G., Dallacasa D., Dolag K., et al., 2010, A\&A, 513, A30. doi:10.1051/0004-6361/200913696
\bibitem[\protect\citeauthoryear{Bonafede et al.}{2022}]{Bonafede2022} Bonafede A., Brunetti G., Rudnick L., Vazza F., Bourdin H., Giovannini G., Shimwell T.~W., et al., 2022, ApJ, 933, 218. doi:10.3847/1538-4357/ac721d
\bibitem[\protect\citeauthoryear{Brunetti et al.}{2001}]{Brunetti2001} Brunetti G., Setti G., Feretti L., Giovannini G., 2001, MNRAS, 320, 365. doi:10.1046/j.1365-8711.2001.03978.x
\bibitem[\protect\citeauthoryear{{Brunetti} \& {Jones}}{{Brunetti} \& {Jones}}{2014}]{BrunettiJones14} {Brunetti} G.,  {Jones} T.~W.,  2014, IJMP D, 23, 1430007. doi:10.1142/S0218271814300079
\bibitem[\protect\citeauthoryear{Brunetti, Zimmer, \& Zandanel}{2017}]{BrunettiZimmer2017} Brunetti G., Zimmer S., Zandanel F., 2017, MNRAS, 472, 1506. doi:10.1093/mnras/stx2092
\bibitem[\protect\citeauthoryear{Cavaliere \& Fusco-Femiano}{1978}]{Cavaliere1978} Cavaliere A., Fusco-Femiano R., 1978, A\&A, 70, 677
\bibitem[\protect\citeauthoryear{Dennison}{1980}]{Dennison1980} Dennison B., 1980, ApJL, 239, L93. doi:10.1086/183300
\bibitem[\protect\citeauthoryear{{Gitti}, {Brunetti}  \& {Setti}}{{Gitti} et~al.}{2002}]{GittiEtAl02} {Gitti} M.,  {Brunetti} G.,   {Setti} G.,  2002, AAP, 386, 456. doi:10.1051/0004-6361:20020284
\bibitem[\protect\citeauthoryear{{Hou}, {Hallinan}  \& {Keshet}}{{Hou} et~al.}{2023}]{HouEtAl22} {Hou} K.-C.,  {Hallinan} G.,   {Keshet} U.,  2023, MNRAS, 521, 5786. doi:10.1093/mnras/stad785
\bibitem[{{Ilani} {et~al.}(2024a){Ilani}, {Hou}, \& {Keshet}}]{IlaniEtAl24}{Ilani}, G., {Hou}, K.-C., \& {Keshet}, U. 2024, arXiv:2402.16946. doi:10.48550/arXiv.2402.16946
\bibitem[{{Ilani} {et~al.}(2024b){Ilani}, {Hou}, {Nadler}, \& {Keshet}}]{IlaniEtAl24b}{Ilani}, G., {Hou}, K.-C., {Nadler}, G., \& {Keshet}, U. 2024, arXiv:2402.17822. doi:10.48550/arXiv.2402.17822
\bibitem[\protect\citeauthoryear{Kamae et al.}{2006}]{Kamae2006} Kamae T., Karlsson N., Mizuno T., Abe T., Koi T., 2006, ApJ, 647, 692. doi:10.1086/505189
\bibitem[\protect\citeauthoryear{Katz \& Waxman}{2008}]{Katz2008} Katz B., Waxman E., 2008, JCAP, 2008, 018. doi:10.1088/1475-7516/2008/01/018
\bibitem[Keshet(2010)]{Keshet10} Keshet, U.\ 2010, arXiv:1011.0729. doi:10.48550/arXiv.1011.0729
\bibitem[{{Keshet} \& {Reiss}(2018)}]{KeshetReiss18}{Keshet}, U. \& {Reiss}, I. 2018, \apj, 869, 53
\bibitem[Keshet(2023)]{Keshet23} Keshet, U.\ 2023, MNRAS, 527, 1194. doi:10.1093/mnras/stad3154
\bibitem[\protect\citeauthoryear{{Keshet} \& {Loeb}}{{Keshet} \& {Loeb}}{2010}]{KeshetLoeb10} {Keshet} U.,  {Loeb} A.,  2010, ApJ, 722, 737. doi:10.1088/0004-637X/722/1/737 722, 737
\bibitem[\protect\citeauthoryear{{Keshet}, {Kushnir}, {Loeb}  \& {Waxman}}{{Keshet} et~al.}{2017}]{KeshetEtAl12_Coma} {Keshet} U.,  {Kushnir} D.,  {Loeb} A.,   {Waxman} E.,  2017, ApJ, 845, 24. doi:10.3847/1538-4357/aa794b
\bibitem[\protect\citeauthoryear{Kushnir \& Waxman}{2009}]{Kushnir2009} Kushnir D., Waxman E., 2009, JCAP, 2009, 002. doi:10.1088/1475-7516/2009/08/002
\bibitem[\protect\citeauthoryear{Kushnir, Katz, \& Waxman}{2009}]{Kushnir2009b} Kushnir D., Katz B., Waxman E., 2009, JCAP, 2009, 024. doi:10.1088/1475-7516/2009/09/024
\bibitem[\protect\citeauthoryear{{Reiss} \& {Keshet}}{{Reiss} \& {Keshet}}{2018}]{ReissKeshet18} {Reiss} I.,  {Keshet} U.,  2018, JCAP, 010, doi:10.1088/1475-7516/2018/10/010
\bibitem[\protect\citeauthoryear{Reiprich \& B{\"o}hringer}{2002}]{Reiprich2002} Reiprich T.~H., B{\"o}hringer H., 2002, ApJ, 567, 716. doi:10.1086/338753
\bibitem[\protect\citeauthoryear{van Weeren et al.}{2019}]{VanWeeren2019} van Weeren R.~J., de Gasperin F., Akamatsu H., Br{\"u}ggen M., Feretti L., Kang H., Stroe A., et al., 2019, SSRv, 215, 16. doi:10.1007/s11214-019-0584-z
\bibitem[\protect\citeauthoryear{Xi et al.}{2018}]{Xi2018} Xi S.-Q., Wang X.-Y., Liang Y.-F., Peng F.-K., Yang R.-Z., Liu R.-Y., 2018, PhRvD, 98, 063006. doi:10.1103/PhysRevD.98.063006
\end{thebibliography}

% Don't change these lines
\bsp	% typesetting comment
\label{lastpage}
\end{document}